\documentclass[twocolumn, prx, reprint, amsfonts, amsmath, amssymb, superscriptaddress, longbibliography, floatfix,aps]{revtex4-1}
\usepackage[T1]{fontenc}
\usepackage{lmodern}
\usepackage{graphicx}
\usepackage[dvipsnames]{xcolor}
\usepackage[colorlinks,linkcolor=Blue,citecolor=Blue]{hyperref}
\usepackage{braket}
\usepackage{mathtools}
\usepackage{enumitem}
\usepackage{microtype}

\begin{document}

\title{High-Temperature Quantum Oscillations of a Non-equilibrium Non-Fermi Liquid}

\author{Oles Matsyshyn}
\affiliation{Division of Physics and Applied Physics, School of Physical and Mathematical Sciences, Nanyang Technological University, Singapore 637371}

\author{Li-kun Shi}
\affiliation{Center for Quantum Matter, School of Physics, Zhejiang University, Hangzhou 310058, China}

\author{Inti Sodemann Villadiego}
\affiliation{Institut f{\"u}r Theoretische Physik, Universit{\"a}t Leipzig, Br{\"u}derstra{\ss}e 16, 04103, Leipzig, Germany}

\date{\today}

\begin{abstract}
A periodically driven Fermi gas coupled to a simple boson bath reaches a non-equilibrium steady-state occupation with sharp non-analyticities at certain momenta. Here, we demonstrate that these non-analyticities behave as emergent Fermi surfaces by showing that they give rise to quantum oscillations of observables with a period controlled by the effective Fermi surface area enclosed by these non-analyticities. However, these oscillations have several striking differences with standard equilibrium quantum oscillations. For example, they remain non-analytic at finite temperatures, their amplitude can survive up to extremely high temperatures comparable to the frequency of the drive, and they can display non-monotonic temperature dependence completely at odds with standard Lifshits-Kosevich behavior.
\end{abstract}

\maketitle

\section{Introduction.}
\begin{figure}[t]
    \centering
    \includegraphics[width=0.48\textwidth]{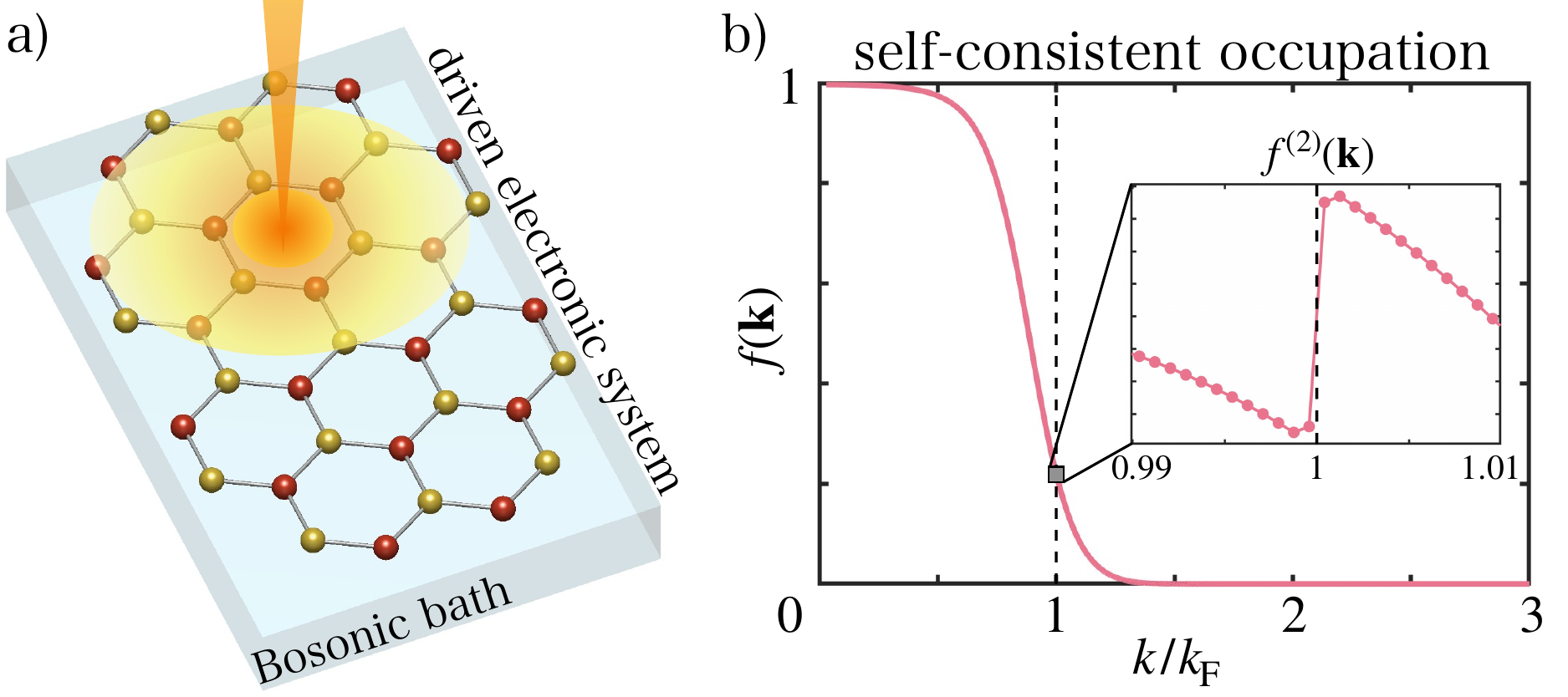}
    \caption{(a) Strongly driven electrons coupled to an ohmic bosonic bath and irradiated by light of frequency $\Omega$. (b) The resulting steady-state occupation develops a discontinuity in its second derivative at momentum $k_F^2/(2m)=\hbar\Omega$. Here we choose $\bar n_e=n_eh/(m\Omega)=0.8$ and $\kappa^2=e^2|\mathcal E|^2/(m\hbar\Omega^3)=0.04$.}
    \label{fig1}
\end{figure}

The Fermi surface is typically viewed as a characteristic of equilibrium Fermi liquids which is sharp only at zero temperature. However, recent studies have demonstrated that the steady state occupations of periodically driven Fermions can display non-analyticities that behave as emergent Fermi surfaces \cite{matsyshyn2023fermi,shi2024floquet,PhysRevLett.134.196401}. In particular, fermions driven periodically with a frequency $\Omega$ (e.g. by monochromatic light) and coupled to a local bosonic ohmic bath, display a jump of the second derivative at wave-vector $k_F^2/2m= \hbar \Omega$ \cite{PhysRevLett.134.196401} (see Fig.\ref{fig1}), which remains sharp even when the bath is at finite temperature and gives rise to phenomena reminiscent to those of equilibrium non-Fermi liquids  \cite{PhysRevLett.134.196401}. 

In this study, we will demonstrate that these non-analyticities behave as emergent Fermi surfaces by showing that they give rise to quantum oscillations in response to magnetic fields with a period controlled by their enclosed area in momentum space. However, we will show that many of their properties differ sharply from those of equilibrium Fermi liquids, further substantiating the idea that they behave like the Fermi surface of a non-equilibrium non-Fermi liquid state. For example, we will demonstrate that, amazingly, the quantum oscillations of these non-equilibrium states are non-analytic as a function of magnetic field even at finite temperature, and that their amplitude is a non-monotonic function of temperature that can survive up to extremely high temperatures comparable to the driving frequency, which is completely at odds with equilibrium behavior \cite{shoenberg1984magnetic}.

\section{Periodically driven fermions coupled to a boson bath.}
We consider a system of fermions driven by a time-periodic Hamiltonian $\hat{H}_{\rm e}(t) = \hat{H}_{\rm e}(t+T)$. This system is coupled via $\hat{H}_{\rm e-b}$ to a bosonic bath governed by $\hat{H}_{\rm b}$, yielding the total Hamiltonian:
\begin{equation}
\hat{H}(t) = \hat{H}_{\rm e}(t) + \hat{H}_{\rm b} + \hat{H}_{\rm e-b},
\label{Full-Hamiltonian}
\end{equation}
where $\hat{H}_{\rm e}(t) = \sum_{\alpha \beta} \bra{\alpha} \hat{h}_{t} \ket{\beta}  \hat{c}_\alpha^\dagger \hat{c}_\beta$, $\hat{H}_{\rm b} = \sum_q \hbar \omega_q \hat{b}^\dagger_q \hat{b}_q$, and $\hat{H}_{{\rm e-b}} = \sum_{q,\nu \eta} \bra{\nu} \hat{\chi}_{q} \ket{\eta} \hat{b}_q \hat{c}_\nu^\dagger \hat{c}_\eta  + \text{h.c.}$. Here, $\hat{h}_t$ is the single-particle system Hamiltonian, and $\hat{\chi}_{q}$ describes the coupling to bath mode $q$. The operators $\hat{c}_\alpha^\dagger$, $\hat{c}_\alpha$ ($\hat{b}_q^\dagger$, $\hat{b}_q$) create and annihilate fermions (bosons) in state $\ket{\alpha}$ (mode $q$). The bath is assumed to be in thermal equilibrium, with Bose-Einstein occupation $N_q = 1/(e^{\beta_T \hbar\omega_q} - 1)$ and temperature $k_B T_{\rm bath} = 1/\beta_T$. The electron system, however, can be driven far from equilibrium through the periodic drive in $\hat{H}_{\rm e}(t)$. Starting from the full quantum kinetic equation discussed in Refs.\cite{VaskoRaichev,PhysRevLett.134.196401}, for the system one-body density matrix, $\rho_{\nu\eta}=\langle c^\dagger_\eta c_\nu \rangle$, which contains memory effects, one can prove that in the limit of weak system-bath coupling ($\chi_q \rightarrow 0$), the steady state of this matrix has Periodic Gibbs ensemble form~\cite{lazarides2014periodic,lazarides2014equilibrium,lazarides2015fate,khemani2016phase}:
\begin{align}
\begin{aligned}
\hat\rho_t &= \sum_{\alpha} f_\alpha \ket{\psi^F_\alpha(t)} \bra{\psi^F_\alpha(t)},
\end{aligned}
\label{rhot}
\end{align}
where $\ket{\psi^F_\alpha(t)} = e^{-i\epsilon^F_\alpha t/\hbar} \sum_{n=-\infty}^{\infty} e^{-in\Omega t} \ket{\phi^n_{\alpha}}$ are the solutions of the time-dependent single particle Schrödinger equation for $h_t$, $\epsilon^F_\alpha$ are the Floquet energies, and $\Omega = 2\pi/T$. {Hereafter, all summations over Floquet modes (for example, those labeled by $n$) are understood to run from $-\infty$ to $\infty$, and we suppress these limits for brevity.} This solution reflects the synchronization of the system with the external periodic drive. The occupations of the Floquet states $f_\alpha$, are time-independent and are obtained by solving the following Floquet-Boltzmann equation [see Appendix \ref{appA} of the Supplemental Material~\cite{SM}]:
\begin{equation}
\sum_\beta f_\alpha \bar{f}_\beta \langle W^{\alpha \to \beta}_t \rangle_T - \sum_\beta \bar{f}_\alpha f_\beta \langle W^{\beta \to \alpha}_t \rangle_T = 0,
\label{FBz}
\end{equation}
where $\langle W^{\alpha \to \beta}_t \rangle \equiv \int_0^T  (W^{\alpha \to \beta}_{e,t} + W^{\alpha \to \beta}_{a,t}) dt/T$ is the time averaged transition rate from state $\alpha$ to $\beta$ and  $\bar{f}_\alpha = 1 - f_\alpha$. The emission rate is given by:
\begin{align}
\begin{aligned}
&\langle W^{\alpha \to \beta}_{e,t} \rangle \hspace{-1mm}=\hspace{-1mm} \frac{2\pi}{\hbar}\hspace{-0.75mm} \sum_{n,q} \hspace{-0.5mm}|V^{n}_{\alpha \to \beta}(q)|^2 \delta(\epsilon^F_\beta \hspace{-0.25mm}-\hspace{-0.25mm} \epsilon_\alpha^F \hspace{-0.25mm}+\hspace{-0.25mm} \hbar\omega_q \hspace{-0.25mm}+\hspace{-0.25mm} n\hbar\Omega), \\
&|V^{n}_{\alpha \to \beta}(q)|^2 \hspace{-1mm}=\hspace{-1mm} (N_q + 1) \left| \sum_m \bra{\phi^m_{\alpha}} \hat{\chi}_q \ket{\phi^{m+n}_{\beta}} \right|^2.
\end{aligned}
\label{Wet}
\end{align}
while the absorption rate $\langle W^{i \to f}_{a,t} \rangle_T$ is obtained from Eq.(\ref{Wet}) by substitution $N_{q}+1\to N_{q},~ \omega_q \to-\omega_q,~\hat\chi\to\hat\chi^\dagger$ \cite{VaskoRaichev}. The above results generalize the results of Ref.\cite{PhysRevLett.134.196401} to arbitrary periodically driven free fermion Hamiltonians with an arbitrary vertex coupling to a boson bath. Similar Floquet Boltzmann equations have also been derived in Refs.\cite{seetharam2015controlled,genske2015floquet,esin2018quantized,seetharam2019steady}. Notice that the above scattering rates obey a generalized form of Fermi's Golden rule. This rule allows violations of Floquet energy conservation modulo $\Omega$ (Floquet umklapp). Crucially, these rates do not obey detailed balance. In the limit of vanishing amplitude of the drive, the Floquet wavefunction becomes static and \( |\phi_{\alpha,n}\rangle \rightarrow \delta_{n0}\ket{\phi_{\alpha,0}} \), the \textit{Floquet Fermi’s Golden Rule} reduces to the standard Fermi’s Golden Rule.

\begin{figure*}
\centering
\includegraphics[width=0.95\textwidth]{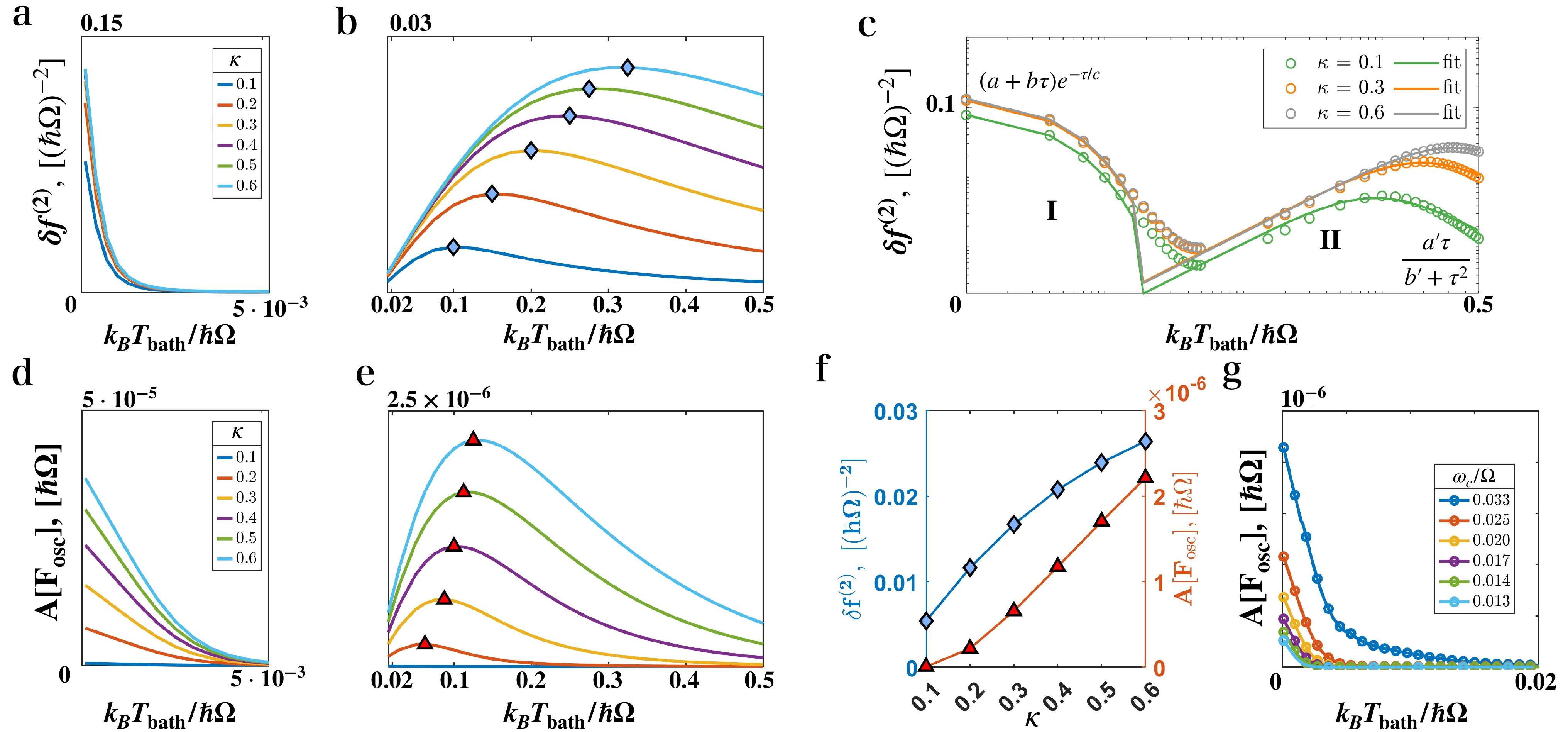}
\caption{Jump of the second derivative of the steady-state occupation and the resulting quantum-oscillation amplitude. Top row: temperature dependence of the jump $\delta f^{(2)}$ for $\bar n_e=0.8, c_2=1/2$ and several driving amplitudes $\kappa$, shown in (a) the low-temperature regime, (b) the high-temperature regime, and (c) on a log-log scale over the full temperature range together with the fits discussed in Sec.~\ref{sec3}. Bottom row: amplitude of the resulting quantum oscillations in the interval $\Omega/\omega_c\in[40,43]$ in (d) the low-temperature and (e) the high-temperature regimes. (f) Comparison between the jump $\delta f^{(2)}$ and the oscillation amplitude maxima as a function of driving amplitude in the high temperature regimes. (g) Low-temperature Floquet oscillations for different magnetic fields, showing that the full width at half maximum (FWHM) scales approximately linearly with $\omega_c$.}
\label{fig3}
\end{figure*}

\section{Floquet non-Fermi liquid at zero magnetic field.}\label{sec3}
As demonstrated in Ref.\cite{PhysRevLett.134.196401}, the steady-state of Fermions driven by monochromatic light and coupled to boson baths displays emergent non-analyticities at certain momenta which resemble the Fermi surfaces of a non-Fermi liquid. Let us review the nature of these states before generalizing to finite magnetic fields. We take a system of parabolic fermions subjected to uniform monochromatic radiation:
\begin{align}
\begin{aligned}
\hat h_t &= \left[\mathbf{p}-e\mathbf{A}(t)\right]^2/(2m),
\end{aligned}
\label{Ht0}
\end{align}
where $\mathbf{A}(t) = -i\Omega^{-1}\boldsymbol{\mathcal{E}}_+ \exp(-i\Omega t)+c.c.$ with $\boldsymbol{\mathcal{E}}_-=(\boldsymbol{\mathcal{E}}_+)^*$. The intensity of the light will be parametrized by the dimensionless constant ${\kappa^2} = e^2|\mathcal{E}|^2/(m\hbar\Omega^3)$. The bath is taken as a collection of harmonic oscillators coupled to the on-site electron density. Namely, the boson modes in Eq.\eqref{Full-Hamiltonian} are labeled as $q=({\bf R},\lambda)$, where $\lambda$ counts the different oscillators at site ${\bf R}$. All modes are coupled with the same strength, $\chi_0$, to the fermions, namely:  $\hat{\chi}_{q} = \chi_0 \ket{{\bf R}} \bra{{\bf R}}$. We take the oscillator frequencies to approach a continuum density of states (DOS)  given by $\nu_B (\epsilon) \propto ( \epsilon + c_2 \epsilon^2) \Theta(\epsilon)$, where $c_2$ parametrizes the deviation from ideal ohmic bath (realized when $c_2=0)$. 

Because of translational invariance, the steady state occupation obtained by solving Eq.(\ref{FBz}) is a function of momentum, $f_\alpha \rightarrow f({\bf k})$. As shown in Ref.\cite{PhysRevLett.134.196401}, for an ohmic bath this steady state displays discontinuities of the second derivative of $f(\mathbf{k})$ at $\mathbf{k}_F^2/(2m)= n\hbar\Omega$ for integer $n$, which behave as the Fermi surface of a non-Fermi liquid state \cite{foot1}\nocite{giamarchi2003quantum,senthil2008critical,varma2002singular}. For concreteness we will focus on the Fermi surface at $n=1$ which is the strongest at small driving amplitudes. Notice that the effective Fermi energy is controlled by driving frequency $\Omega$, unlike the equilibrium Fermi energy which is determined by the fermion density. Figure \ref{fig3} (a-c)  shows the behavior of the second order discontinuity $\delta f^{(2)}\equiv f''[\epsilon = \hbar\Omega + 0]-f''[\epsilon = \hbar\Omega - 0]$ for the $n=1$ Fermi surface obtained directly from numerical solutions of Eq.(\ref{FBz}) (see also Fig.1(b)), corroborating the remarkable finding of Ref.\cite{PhysRevLett.134.196401} that these non-analyticities remain sharp at finite bath temperature (i.e. the ``ultra-critical'' behavior).

The strength of this emergent Fermi surface (i.e. the size of the discontinuity) is non-monotonic as a function of the bath temperature, as shown in Fig.~\ref{fig3} (a)-(c). At low temperatures ($k_BT_{\rm bath} \ll \hbar\Omega$), the strength of the emergent Fermi surface first decreases rapidly with temperature and can be fit as $\delta f^{(2)}[\tau \rightarrow 0] = (a + b\tau)e^{-\tau/c},$ where \( \tau = k_B T_{\rm bath} / \hbar \Omega \) is dimensionless temperature [see Fig.~\ref{fig3}c] [for the fit parameters see Appendix \ref{fit_param}]. The size of the discontinuity in this regime saturates at  large dimensionless light intensity,  \( \kappa^2 \), at values of order $\delta f^{(2)}\sim (\hbar\Omega)^{-2}$.

Remarkably, at higher temperatures there is a revival of the strength of the emergent Fermi surface that can peak at a temperature comparable to frequency of the drive, i.e. $k_BT_{\rm bath} \sim \kappa \hbar\Omega$ as seen in Fig.~\ref{fig3}(b),(f). Amazingly, this implies that the Fermi surface remains sharp at high temperatures comparable to the effective Fermi energy, and its decay at high temperatures can be approximated as a power-law: $\delta f^{(2)}[\tau \gg \tau_{\rm max}] = a'\tau /(b'+\tau^{2})$ [see Fig.~\ref{fig3}c] [for the fit parameters see Appendix \ref{fit_param}]. Strikingly, as we will show in the next section, even such a seemingly weak non-analyticity of the occupation gives rise to quantum oscillations comparable in magnitude to those in an ordinary Fermi liquid. The presence of this higher temperature revival of the Fermi surface is ultimately controlled by the deviation from the perfectly ohmic bath, namely, for $c_2\rightarrow0$ the high temperature revival  is absent [see Appendix \ref{ohmic_bath_osc} of the Supplemental Material~\cite{SM}]. 

\section{Kinetics of driven Landau levels.}
We will now discuss the response of the Floquet non-Fermi liquid to a magnetic field. The electron Hamiltonian from Eq.(\ref{Ht0}) is therefore modified to be:
\begin{align}
\begin{aligned}
\hat h_t &= \left[\mathbf{p}-e\mathbf{A}_0(\mathbf{r})-e\mathbf{A}(t)\right]^2/(2m), 
\end{aligned}
\label{Ht}
\end{align}
where $\boldsymbol{\nabla}\times \mathbf{A}_0(\mathbf{r}) = \mathbf{B}_0$ accounts for the constant magnetic field. Solutions to Schrödinger's associated with Eq.~(\ref{Ht}) have been discussed in quantum optics \cite{osti_5185075,Gerry_Knight_2004}, and can be obtained via the unitary  transformation $D[\alpha(t)] = \exp[\alpha(t) \hat{a}^\dagger - \alpha^*(t) \hat{a}]$ where $\alpha(t)=\sum_{\xi=\pm1}z_\xi e^{-i\xi\Omega t}/(\hbar\omega_c-\xi\hbar\Omega)$ with $z_\xi=-i \xi \Omega^{-1}\sqrt{\hbar\omega_c/(2m)}(\mathcal{E}_\xi^x-i\mathcal{E}^y_\xi)$, as follows [see Appendix \ref{non_pert} of the Supplemental Material~\cite{SM} and Ref.\cite{shi2025algebraic} for more details]:
\begin{align}
\begin{aligned}
\ket{N,r,t} &= e^{-i\Phi_N(t)} D[\alpha(t)] \ket{N} \otimes \ket{r}, \\&= e^{-i\Phi_N(t)} \sum_{n=-\infty}^\infty e^{-in\Omega t} \ket{\phi^{n}_{N}} \otimes \ket{r},
\end{aligned}
\label{state}
\end{align}
where $\Phi_N(t) = E_N(t - t_0)/\hbar + (1/\hbar)\int_{t_0}^t \Delta(t')\,dt'$, $E_N = \hbar\omega_c(N + 1/2)$, $N$ is the Landau level index, $r$ is the intra-Landau level guiding center index, $\omega_c= e |\mathbf{B}_0|/m$ is the cyclotron frequency and $\Delta(t)$ is a periodic, Landau-level-index-independent scalar energy shift generated by the displacement transformation; its explicit form is given in Appendix \ref{non_pert} of the Supplemental Material~\cite{SM}. In the second line of Eq.(\ref{state}) we used the Floquet theorem to express the wave-function as a sum of Floquet harmonics. 

Owing to magnetic translational symmetry, the scattering rates and occupations become independent of the guiding center index $r$ [see Appendix \ref{guiding_center} of the Supplemental Material~\cite{SM}], allowing Eq.~(\ref{Wet}) to be recast as:
\begin{align}
\begin{aligned}
&\langle{W}^{N\rightarrow N'}_{e,t}\hspace{-0.5mm}\rangle_{T}\hspace{-1mm}=\hspace{-1mm} \frac{2\pi}{\hbar}\hspace{-1mm}\sum_{n}\hspace{-1mm}\int_{-\infty}^\infty \hspace{-4.2mm} d\epsilon ~\nu_B(\epsilon)|\hspace{-0.25mm}V^{n}_{N N'}(\epsilon)\hspace{-0.25mm}|^2\hspace{-0.5mm}\delta( E^n_{N'N}\hspace{-0.75mm}+\hspace{-0.75mm}\epsilon),\\&|V^{n}_{N N'}(\epsilon)|^2= \frac{\chi_0^2}{4\pi^2l_B^4}(N_b(\epsilon)+1)Q^n_{NN'}\\&Q^n_{NN'}\hspace{-1mm}=\hspace{-1mm}{\rm Tr}\hspace{-0.5mm}\left\{\hspace{-1mm}\bigg[\hspace{-1mm}\sum_m \ket{\phi^{m}_{N}}\bra{\phi^{m-n}_{N'}}\hspace{-0.75mm}\bigg]\hspace{-1mm}\bigg[\hspace{-1mm}\sum_{m'}\ket{\phi^{m'}_{N}}\bra{\phi^{m'-n}_{N'}}\hspace{-0.75mm}\bigg]^\dagger\hspace{-0.5mm}\right\},
\end{aligned}\label{WetLL}
\end{align}
where $E^{n}_{NN'} = E_N - E_{N'}+n\hbar\Omega$, and $l_B$ is the magnetic length. The trace can be evaluated with the following identity:
\begin{multline}
    Q^n_{NN'}=
    \iint_0^1 dx\, dy\, e^{i 2\pi n (x-y)}\\
    \times e^{-|\beta(x,y)|^2}
    L_{N'}\left(|\beta(x,y)|^2\right)
    L_N\left(|\beta(x,y)|^2\right),\label{genTR}
\end{multline}
where $L_N(x)$ is the Laguerre polynomial. For circularly polarized light: $|\beta(x,y)|^2 = 2\kappa^2(\omega_c/\Omega) [1 - \cos(2\pi(x-y))]$ [see Appendix \ref{trace_eval} of the Supplemental Material~\cite{SM} for more general form]. In the limit $\kappa \to 0$, Eq.~(\ref{genTR}) vanishes for $n\neq0$, and Eq.~\eqref{FBz} reduces to the conventional Boltzmann equation. However, the above equations are non-perturbative and valid for any value of $\kappa$. Equipped with Eq.~(\ref{genTR}) and Eq.~(\ref{WetLL}), we numerically solve the Floquet-Boltzmann Eq~\eqref{FBz}.
\begin{figure}[t]
    \centering
    \includegraphics[width=0.49\textwidth]{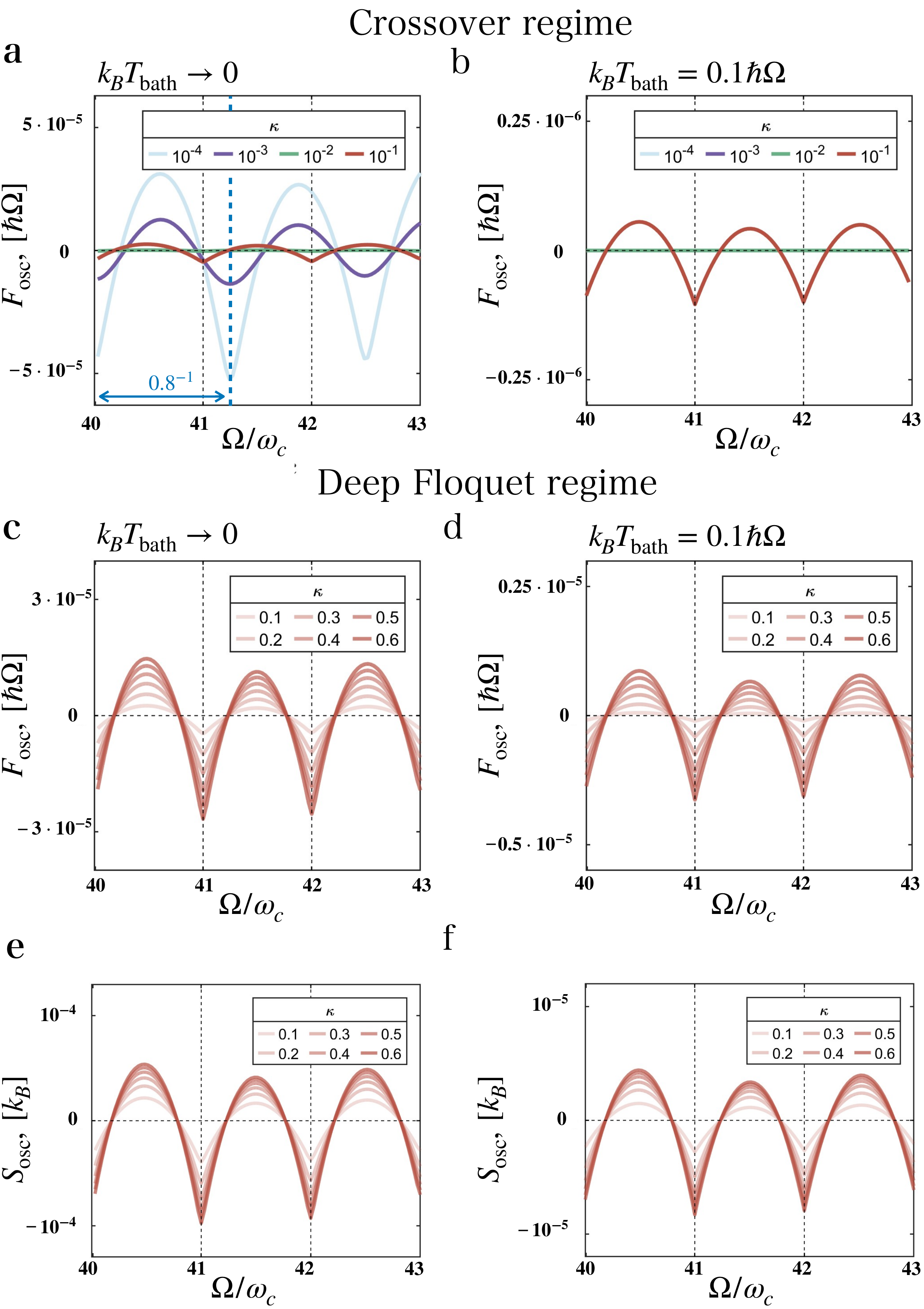}
    \caption{(a) Quantum oscillations at small driving amplitudes $\kappa$ and low bath temperature. In the limit $\kappa \to 0$, only equilibrium oscillations, controlled by the particle density, are visible; these are rapidly suppressed as $\kappa$ increases, giving way to emergent Floquet oscillations with a period dictated by $\Omega$. (b) At higher bath temperatures, equilibrium oscillations are exponentially suppressed, while Floquet oscillations remain distinct and non-analytic as a function of magnetic field. (c,d) At larger driving amplitudes, only Floquet quantum oscillations persist in both the low- and high-temperature regimes. (e,f) Corresponding low- and high-temperature quantum oscillations of the von Neumann entropy.}
   \label{fig2}
\end{figure}

\smallskip

\section{Ultra-critical quantum oscillations.}
We now illustrate that the non-analyticity of the occupation indeed behaves as the Fermi surface of a non-Fermi liquid by showing that it gives rise to quantum oscillations with markedly different behavior compared to an equilibrium Fermi liquid. Quantum oscillations are expected to appear in many observables, but for concreteness we will focus on a non-equilibrium free energy defined as:
\begin{equation}\label{F_full}
    \frac{F}{n_e l_B^2 } \equiv \frac{1}{n_e l_B^2}\sum_{N=0}^\infty\left[ f_N E_N  - T_{\rm bath}  S_N\right],
\end{equation}
where $S_N = - k_B(f_N \log f_N + \bar f_N \log \bar f_N)$ is the entropy contribution associated with the filling $f_N$ [see Appendix \ref{appB} of the Supplemental Material~\cite{SM}]. This free energy can be obtained from the physical energy of the system averaged over one period [see Appendix \ref{appB} of the Supplemental Material~\cite{SM} for detailed derivation] and subtracting the entropy multiplied by the bath temperature. Therefore, this free energy is a linear combination of two physically meaningful quantities (the average energy and the entropy) chosen such that in the limit of weak driving ($\kappa\rightarrow0$) it approaches the standard equilibrium Helmholtz free energy, which is the natural thermodynamic potential when a system is coupled to a bath with which it only exchanges energy. However, some caution should be exercised in interpreting this non-equilibrium free energy, because equilibrium relations such as those allowing to compute other physical quantities (like heat capacity or magnetization) from its derivatives do not necessarily hold away from equilibrium.

We have evaluated Eq.(\ref{F_full}) by numerically solving the Boltzmann Eq.(\ref{FBz}) using scattering rates that are perturbative and non-perturbative in $\kappa$ [see Appendix \ref{trace_eval} of the Supplemental Material~\cite{SM}], and verified that both calculations give the same behavior at small $\kappa$ [see Appendix \ref{trace_eval} of the Supplemental Material~\cite{SM}]. The perturbative rates are more efficient for numerical simulations, and we used them to explore in more detail the full temperature dependence. 

Figure \ref{fig2} shows that oscillations with a period associated with an effective Fermi energy $\hbar \Omega$ emerge as $\kappa$ increases. These oscillations can be easily distinguished from the standard equilibrium oscillations which have a different period controlled by electron density, and which decrease as $\kappa$ increases as a result of electron heating. One of our most remarkable findings, seen in Fig.\ref{fig2}(b), is that these new oscillations of the emergent Floquet Fermi surface remain non-analytic as a function of {\it magnetic field} even when the bath is at finite temperature. This is a manifestation of their ``ultra-critical'' nature, namely the fact that the emergent Floquet Fermi surface remains sharp at finite temperature \cite{PhysRevLett.134.196401}. In equilibrium quantum oscillations are non-analytic as a function of field only at zero temperature, and happen when the chemical potential jumps across the cyclotron gaps between Landau levels, which manifests as cusp-like features in the energy vs. filling factor plots (i.e. inverse magnetic field) \cite{shoenberg1984magnetic, abrikosov2017fundamentals}. Therefore, what we see in Fig.\ref{fig2}(b) is a non-equilibrium version of this behavior that amazingly remains sharp even at finite temperatures.

We have also found that the amplitude is non-monotonic as a function of temperature, as shown in Fig.~\ref{fig3}, going first through a regime of rapid decrease with bath temperature, followed by a high temperature revival that eventually peaks at a temperature comparable with the driving frequency $\hbar \Omega$ (i.e. the effective Floquet Fermi energy), and subsequently decaying as $\sim 1/T_{\rm bath}$ at higher temperatures. Therefore, the amplitude of oscillations is well correlated with the strength of the emergent Fermi surface, measured from the size of the jump of the second derivative at zero field, $\delta f^{(2)}$, as shown in Fig.\ref{fig3}.

The amplitude of the Floquet quantum oscillations at low temperatures can be comparable [see Appendix \ref{trace_eval} of the Supplemental Material~\cite{SM}] to the amplitude of the usual equilibrium quantum oscillations for strong light intensities $\kappa^2\sim 1$, but it remains up to about 10$\%$ of it even down weaker intensities $\kappa^2 \sim 0.01$. On the other hand, the high temperature oscillations tend to remain relatively weaker in amplitude than the equilibrium oscillations, i.e. for $\kappa^2=1$ they are about a few percent at their highest. The temperature width of the region of high temperature oscillations is weakly dependent on magnetic field, but the temperature width of the low-temperature oscillations increases linearly with magnetic field, i.e. the latter is more similar to usual quantum oscillations \cite{shoenberg1984magnetic,abrikosov2017fundamentals}. All of the above substantiates the idea that the non-analyticity behaves as an emergent Fermi surface, which governs the observed  oscillatory behavior at small magnetic fields, and that can retain quantum features even up to remarkably elevated temperatures. 

The basic mechanism underlying magnetic oscillations is that, as the magnetic field is varied, Landau levels cross the effective Floquet Fermi surfaces of the Floquet non-Fermi-liquid state, which are set by the driving frequency. Therefore, one expects oscillations not only in this free energy but also in a variety of other observables, in analogy to the equilibrium case. To illustrate this explicitly, we have also computed the von Neumann entropy of the state [see Appendix \ref{appB} of the Supplemental Material~\cite{SM} for details] whose oscillations are shown in Fig.~\ref{fig2}(e,f).

We caution readers that while our current calculations are very valuable as proofs of principle, there are many realistic ingredients that are crucial in order to understand the visibility in experiments of these non-equilibrium quantum oscillations that we have so far neglected, and which we hope to incorporate in future studies. One of the crucial ingredients we have neglected is the effect of Landau level broadening. The reason we have neglected this, is that our work focuses on understanding systematically the non-equilibrium steady state in the limit of weak coupling to the bath, i.e. $\hat\chi_{\mathbf{q}}\rightarrow0$ in Eq.~(\ref{Full-Hamiltonian}). Notice that the non-equilibrium occupations that solve the Boltzmann Eq.~(\ref{FBz}) are of order zero in the system bath coupling. Thus the leading effect in the limit $\hat\chi_{\mathbf{q}}\rightarrow0$ is to re-shuffle the occupation of states by a large amount ultimately controlled by the amplitude of the drive (but approximately independent of the strength of the coupling to the bath). At higher orders in perturbation theory, a finite coupling to the bath gives rise also self-consistently to the finite broadening $\Gamma$ of the Landau levels. 

A systematic account of such broadening requires a self-consistent non-equilibrium Greens function calculation, which is beyond the scope of our work. Nevertheless, we would like to comment on the  broad expectations arising from accounting for this broadening (which for real experiments would also need to account not only for the coupling to the bath, but also for the effects of disorder and electron-electron interactions). In equilibrium quantum-oscillation problems, finite level broadening $\Gamma$ suppresses the oscillation amplitude through the usual Dingle damping factor, e.g. $R_D=\exp[-\pi \Gamma/\omega_c]$ (see e.g. Ref.~\cite{RevModPhys.84.1709}). Thus we expect that also in the present driven problem additional broadening and finite lifetime effects would reduce the oscillation amplitude and eventually wash it out once the broadening becomes comparable to the Landau-level spacing. Characterizing the impact of this broadening in a systematic fashion in the non-equilibrium setting remains an important direction for future work.

\section{Guiding experiments towards true Floquet steady states.}

One major open experimental challenge is to realize {\it true steady state} occupations of Floquet bands and not simply short-lived transients. Currently, a large portion of Floquet experiments is performed with ultra-fast optical techniques. While they have produced many beautiful results, such as the realization of Floquet states in topological insulators surfaces \cite{wang2013observation}, the light-induced Hall effect in graphene \cite{mciver2020light}, these approaches only realize Floquet states for short durations (typically picoseconds), and are far from reaching true steady states, making it difficult to probe and model their behavior. Reaching such steady states would highly facilitate the observation of our non-equilibrium states with emergent Fermi surfaces, because it would allow to perform similar experiments to those employed to detect Fermi surfaces in equilibrium (e.g. quantum oscillations under magnetic fields). How can experiments reach such states? the strength of Floquet effects is typically controlled by the work performed by the electric field during one cycle divided by the photon energy, namely by the
dimensionless parameter:
\begin{equation}
\beta_F = \frac{v\,e\,|\mathcal{E}|}{\hbar\,\Omega^{2}},
\end{equation}
where $v$ is the typical electron velocity. 

Floquet effects start to become visible when the dimensionless light intensity, $\beta_F^{2}$, is not too small. For example, for ultra-fast optical approaches typically performed using mid-infrared light with $\hbar\omega \sim 120\,{\rm meV}$, Floquet effects can be visible when $\beta_F^{2} \sim 0.1$, but this requires extremely high intensities typically of the order $10^{12}\rm Watt/\ m^{2}$. Such high intensities can lead to extreme heating of the samples, and thus are typically applied only on very short picosecond pulses (see however \cite{ref10,wang2013observation}). But notice that $\beta_F^{2}$ not only increases with light intensity, but also very fast as $\Omega^{-4}$ as frequency is lowered. Therefore, we believe this to be a highly promising avenue to reach Floquet states under continuous illumination because it permits to reach large values of $\beta_F^{2}$ at low intensities, leading to less heating and allowing the system to settle into truly quantum non-equilibrium steady states. Ultimately, Floquet effects can be washed out once the frequency becomes comparable to quasi-particle lifetime scales, and thus it is desirable to have ultra-clean materials and push the frequency to be as low as possible. Are there precedents for reaching this regime? Yes, to some extent. A variety of beautiful steady state effects have been observed in 2DEGs like GaAs heterostructures, known as micro-wave induced resistance oscillation (MIRO) phenomena \cite{RevModPhys.84.1709, shi2015shubnikov}. For example, in GaAs with 10 GHz only an intensity of $\rm 0.2{\rm Watt}/\ m^{2}$ is needed to reach $\beta_F^{2} \sim 0.1$. But these investigations have largely not yet expanded to other quantum materials (e.g. only one experiment has reported related phenomena in graphene \cite{ref13}). 

We anticipate that interactions, by enhancing scattering processes, will lead to hotter steady states compared to the non-interacting case. Since the nonanalytic features survive at finite temperature and decay only as a power law, this additional heating should not eliminate them, although their magnitude will be reduced. More importantly, interactions generate a finite quasiparticle lifetime, an effect that lies beyond the Floquet–Boltzmann description. We expect that as long as this lifetime exceeds the Floquet period, the non-Fermi surfaces should remain observable. Clarifying this criterion, as well as the role of disorder and the minimal frequency required for their stability as true steady states are very important questions that we hope to systematically address in future investigations.

\section{Summary.}
We derived a Floquet-Boltzmann kinetic equation for the occupation of states in a periodically driven fermionic system coupled to an ohmic bosonic bath. We demonstrated that its solutions display non-analyticities at certain momenta which behave like emergent Fermi surfaces, by showing that they give rise to quantum oscillations of observables in the presence of magnetic fields. These oscillations have many remarkable differences with respect to equilibrium quantum oscillations, e.g. their period is controlled by the driving frequency (not the electron density), they are non-analytic as a function of magnetic field, non-monotonic as a function of temperature, can survive up to extremely high temperatures comparable to the driving frequency, and their amplitude can be comparable to those in equilibrium.

\section{Acknowledgments.} 
We would like to thank Jurgen Smet, Andrei Pimenov, Alexey Shuvaev, and Michael Zudov for stimulating discussions. We acknowledge support by the Deutsche Forschungsgemeinschaft (DFG) through research grant project numbers 542614019; 518372354; 555335098.

\bibliography{floquet-non-fermi-liquid}
\clearpage
\appendix
\renewcommand{\theequation}{\thesection-\arabic{equation}}
\renewcommand{\thefigure}{\thesection-\arabic{figure}}
\renewcommand{\thetable}{\thesection-\Roman{table}}
\onecolumngrid
\setcounter{page}{1}

\begin{center}
{\large\bfseries Supplemental Material for\\[2mm]
``High-Temperature Quantum Oscillations of a Non-equilibrium Non-Fermi Liquid''}

\vspace{2mm}

Oles Matsyshyn,$^{1}$ Li-kun Shi,$^{2}$ and Inti Sodemann Villadiego$^{3}$

\vspace{2mm}

{\textit{\small
$^{1}$Division of Physics and Applied Physics, School of Physical and Mathematical Sciences,\\
Nanyang Technological University, Singapore 637371\\
$^{2}$Center for Quantum Matter, School of Physics, Zhejiang University, Hangzhou 310058, China\\
$^{3}$Institut f\"{u}r Theoretische Physik, Universit\"{a}t Leipzig, Br\"{u}derstra{\ss}e 16, 04103 Leipzig, Germany
}}
\end{center}

\section{Floquet Boltzmann equation}\label{appA}
In this section we derive the scattering rates entering Eq.(\ref{FBz}) of the main text. Our goal is to show that they take the form of a Floquet generalization of Fermi’s Golden Rule, Eq.(\ref{Wet}), and to clarify the physical meaning of the different terms.
\begin{equation}
    \langle W^{\alpha \to \beta}_{t} \rangle = \frac{2\pi}{\hbar} \sum_{n,q} |V^{n}_{\alpha \to \beta}(q)|^2 \delta(\epsilon^F_\beta- \epsilon_\alpha^F + \hbar\omega_q + n\hbar\Omega)
\end{equation}
This has the structure of a Fermi Golden Rule, but with Floquet-assisted processes where energy conservation holds modulo multiples of the drive frequency. Unlike equilibrium Fermi Golden Rule, these rates do not satisfy detailed balance.

We start the derivation from the emission part of the collision integral for the one-body electron density matrix evolution $\partial_t \hat \rho_t+(i/\hbar)[\hat H_{\rm e}(t),\hat \rho_t] = (1/\hbar)J_e[t,\hat{\rho}_t]+(1/\hbar)J_a[t,\hat{\rho}_t]$ (see Refs.\cite{VaskoRaichev,PhysRevLett.134.196401}):
\begin{equation}
    J_e[t,\hat{\rho}_t] = \frac{1}{\hbar}\sum_q \int_{t_0}^t dt^\prime (N_{q}+1)e^{-i\omega_q (t-t^\prime)}\left[  U_{\rm e}(t,t^\prime)(1-\hat{\rho}_{t'})\chi^{\dagger}_{q}\hat{\rho}_{t'}U_{\rm e}(t^\prime,t),\chi_{q}\right] +h.c.,\label{Jeappr}
\end{equation}
where $i d_t \hat U_{\rm e}(t,t') =\hat H_{\rm e}(t)\hat U_{\rm e}(t,t')$ is the evolution operator of the electronic subsystem, $[\hat A,\hat B]\equiv \hat A\hat B-\hat B\hat A$ is the standard commutator and the absorption contribution follows from the replacements $N_{q}+1\to N_{q},~ \omega_q \to-\omega_q,~\hat\chi_q\to\hat\chi_q^\dagger$ \cite{VaskoRaichev}, with $q$ above to be a label for a boson quantum state. 

Due to the bath coupling we expect a drive-bath stabilized steady state of the system is reached in the form of the periodic Gibbs ensemble form introduced in Eq.(\ref{rhot}) of the main text (see also Refs.~\cite{lazarides2014periodic,lazarides2014equilibrium,lazarides2015fate,khemani2016phase}). Below we show that indeed there is periodic Gibbs ensemble which is a self-consistent solution of the kinetic equations in the weak coupling limit between the system and the bath. By expanding the evolution operator in the Floquet basis as:
\begin{equation}
U_{\rm e}(t,t') = \sum_{\alpha,n,m}
e^{-i\epsilon^F_\alpha (t-t')} e^{-in\Omega t} e^{im\Omega t'}
|\phi^n_\alpha\rangle \langle \phi^m_\alpha|.
\end{equation}

Projecting the kinetic equation onto the diagonal Floquet components and performing standard manipulations (see Ref.~[5]), we obtain the steady state equation as:
\begin{equation}
    0= \frac{1}{\hbar}\sum_{nm} \bra{\phi^{n}_{\alpha}} (J_e + J_a) \ket{\phi^{m}_{\alpha}}e^{i(n-m)\Omega t}=\frac{1}{\hbar}\mathcal{I}_{e,\alpha}+\frac{1}{\hbar}\mathcal{I}_{a,\alpha}.\label{pBFQ}
\end{equation}
which, as we will see shortly is equivalent to Eq.(\ref{FBz}) of the main text.

Due to non-triviality of the next step, we will evaluate the collision part of the Eq.~(\ref{pBFQ}) step by step starting from writing it explicitly:
\begin{multline}
\frac{1}{\hbar}\mathcal{I}_{e,\alpha}(t)=\frac{1}{\hbar}\sum_{nm}\bra{\phi^n_{\alpha}}J_e[t,\rho]\ket{\phi^m_{\alpha}} e^{i(n-m)\Omega t} = \frac{1}{\hbar^2}\sum_{l_1l_2l_3l_4} \sum_q \int_{t_0}^t dt^\prime (N_{q}+1) e^{-il_1\Omega t+il_2 \Omega t^\prime-il_3\Omega t^\prime+il_4 \Omega t}\times\\\times\bigg[\sum_{\beta'}   e^{-i(\hbar\omega_q+\epsilon^F_{\alpha}-\epsilon^F_{\beta '})(t-t^\prime)/\hbar}(1-f_{\alpha})f_{\beta '}\bra{\phi^{l_2}_{\alpha}}\hat \chi^{\dagger}_{q} \ket{\phi^{l_3}_{\beta '}} \bra{\phi^{l_4}_{\beta '}}\hat\chi_{q}\ket{\phi^{l_1}_{\alpha}}\\- \sum_\beta e^{-i(\hbar\omega_q+\epsilon^F_{\beta }-\epsilon^F_{\alpha})(t-t^\prime)/\hbar}(1-f_{\beta })f_{\alpha}\bra{\phi^{l_4}_{\alpha}}\hat\chi_{q}\ket{\phi^{l_1}_{\beta }}\bra{\phi^{l_2}_{\beta }}\hat\chi^{\dagger}_{q} \ket{\phi^{l_3}_{\alpha}} \bigg] +h.c.
\end{multline}
where we used $\delta_{\alpha\beta} = \sum_{l_1l_2} e^{i(l_1-l_2)\Omega t}\langle \phi_{\alpha}^{l_1}\ket{\phi_{\beta}^{l_2}}$. It is useful to introduce $\chi_{q,\alpha\beta}(\ell)\equiv \sum_{l_0}\bra{\phi^{l_0}_\alpha}\hat{\chi}_q\ket{\phi^{l_0+\ell}_\beta}$ relabel modes as $l_1 = l_4 +\ell,~l_2 = l_3 + \ell'$ and relabel dummy parameter $t'-t = \tau$ to cast the above as:
\begin{multline}
\frac{1}{\hbar}\mathcal{I}_{e,\alpha}(t)=\frac{1}{\hbar}\sum_{nm}\bra{\phi^n_{\alpha}}J_e[t,\rho]\ket{\phi^m_{\alpha}} e^{i(n-m)\Omega t} = \frac{1}{\hbar^2}\sum_{\ell\ell'}e^{-i\ell\Omega t+i\ell' \Omega t} \sum_{q,\beta} \int_{t_0-t}^0 d\tau (N_{q}+1) \times\\\times\bigg[   e^{i(\hbar\omega_q+\epsilon^F_{\alpha}-\epsilon^F_{\beta}+\ell'\hbar \Omega)\tau/\hbar}\bar f_{\alpha}f_{\beta } [\chi_{q,\beta\alpha}(\ell')]^\dagger\chi_{q,\beta\alpha}(\ell)- (\alpha\leftrightarrow\beta) \bigg] +h.c.
\end{multline}
After we employ the standard approximation $\int_{t_0}^0 d\tau f(\tau) = \lim_{\eta \rightarrow 0}\int_{-\infty}^0 d\tau e^{\eta \tau /\hbar} f(\tau)$ we obtain the generic time dependent collision integral as:
\begin{equation}
\frac{1}{\hbar}\mathcal{I}_{e,\alpha}(t)=\frac{1}{\hbar}\sum_{\ell\ell'}e^{-i\ell\Omega t+i\ell' \Omega t} \sum_{q,\beta}  (N_{q}+1) \bigg[  \frac{\bar f_{\alpha}f_{\beta } [\chi_{q,\beta\alpha}(\ell')]^\dagger\chi_{q,\beta\alpha}(\ell)}{i(\hbar\omega_q+\epsilon^F_{\alpha}-\epsilon^F_{\beta}+\ell'\hbar \Omega)+\eta}- (\alpha\leftrightarrow\beta) \bigg] +h.c.
\end{equation}

At this stage, the collision integral remains explicitly time dependent through harmonics of the drive. Since the steady-state occupations $f_\alpha$ are time independent, the relevant kinetic equation is obtained from its static component, i.e. the average over one drive period, which corresponds to $\ell=\ell'$. We then simplify this component using $\frac{1}{ix+\eta}+\frac{1}{-ix+\eta}=\frac{2\eta}{x^2+\eta^2}\rightarrow 2\pi\delta(x),$ which yields the emission part of the Floquet--Boltzmann collision integral in the form
\begin{equation}\label{emSI}
   \frac{1}{\hbar}\bar{\mathcal{I}}_{e,\alpha} =\sum_{\beta} \bigg[  \bar f_{\alpha} f_{\beta} \langle W_{t,e}^{\beta\rightarrow\alpha} \rangle_T- (\beta\leftrightarrow \alpha) \bigg] 
\end{equation}
where $\langle W_{t,e}^{\alpha\rightarrow\beta} \rangle_T =(2\pi/\hbar)\sum_q(N_{q}+1)\sum_\ell|\chi_{q,\alpha\beta}(\ell)|^2\delta(\hbar\omega_q+\epsilon^F_{\beta}-\epsilon^F_{\alpha}+\ell\hbar\Omega)$. The derived result combined with the absorption part is the Eq.(\ref{Wet}) of the main text ($ W_{t}^{\beta\rightarrow\alpha}\equiv  W_{t,e}^{\beta\rightarrow\alpha} +  W_{t,a}^{\beta\rightarrow\alpha}$), has the structure of a Fermi Golden Rule, but generalized to periodically driven systems. Eq.(\ref{FBz}) is obtained from the stationary condition as in Eq.(\ref{pBFQ}). 

Importantly:
\begin{itemize}
\item The integer $n$ labels the exchange of $n$ quanta of the drive with frequency $\Omega$.
\item Energy conservation is modified to $\epsilon^F_\beta - \epsilon^F_\alpha + \hbar\omega_q + n\hbar\Omega = 0$ (Floquet umklapp processes).
\item Absence of detailed balance. This ultimately arises because the Floquet harmonics of a state $\alpha$ are not determined exclusively by its Floquet energy $\epsilon^F_\alpha$. As a result, the key condition needed to establish detailed balance is not satisfied.
\item The matrix element $\langle W_{t}^{\alpha\rightarrow\beta} \rangle_T$ contains overlaps between different Floquet harmonics.
\end{itemize}

In the limit of vanishing drive amplitude, only the $n=0$ term survives and the expression reduces to the standard equilibrium Fermi Golden Rule. A crucial difference with equilibrium is that the rates do not satisfy detailed balance. As a result, the steady state occupations $f_\alpha$ are not constrained to take a Fermi–Dirac form. This lack of detailed balance is ultimately responsible for the non-equilibrium steady states discussed in the main text, including the emergence of non-analytic features in the distribution function.

\section{Floquet spectrum of periodically driven Landau levels}\label{non_pert}
Here we discuss the details of the kinetics of driven Landau levels. First, we rewrite the Hamiltonian in Eq.(\ref{Ht}) in its operator form as:
\begin{align}
\begin{aligned}
\hat h_t &= \left[\hat{\mathbf{p}}-e\mathbf{A}_0-e\mathbf{A}(t)\right]^2/(2m),= \hbar\omega_c\left(1/2+ \hat{a}^\dagger \hat{a}\right) - \hat{a} z_t^* - \hat{a}^\dagger z_t + c_t,
\end{aligned}
\label{HtSI}
\end{align}
where 
\begin{gather}
     z_t = \sqrt{\frac{\hbar\omega_c}{2m}}(A^x(t)-iA^y(t))=z_+ e^{-i\Omega t}+ z_-e^{i\Omega t},\qquad
    {\bf A} (t) = -\frac{i}{\Omega} \boldsymbol{\mathcal{E}}_+ \exp (-i \Omega t) + \frac{i}{\Omega} \boldsymbol{\mathcal{E}}_{-} \exp (i \Omega t),\label{B1}\\
    \mathbf{B} = \nabla \times A_0,\qquad a^\dagger\ket{N} = \sqrt{N+1}\ket{N+1},\qquad a \ket{N} = \sqrt{N}\ket{N-1},\\
    z_+ = -\frac{i}{\Omega}\sqrt{\frac{\hbar\omega_c}{2m}}(\mathcal{E}_+^x-i\mathcal{E}^y_+),\qquad z_- = \frac{i}{\Omega}\sqrt{\frac{\hbar\omega_c}{2m}}(\mathcal{E}_{-}^x-i\mathcal{E}^y_{-}),\qquad
    c_t = \frac{e^2{\bf A}(t)^2}{2m\hbar^2}.\label{B3}
\end{gather}

We proceed by performing the unitary displacement transformation $D = \exp[\alpha a^\dagger -\alpha^* a]$ with $D^\dagger = D^{-1}$~\cite{shi2025algebraic}. In order to explicitly show the displacement transformation, we will employ two useful identities. First is the special form of the Baker–Campbell–Hausdorff formula known as Campbell identity, which explicitly reads as
\begin{equation}
    e^{X}Y e^{-X} = Y +[X,Y] +\frac{1}{2!}[X,[X,Y]]+\frac{1}{3!}[X,[X,[X,Y]]]+\ldots
\end{equation}
which in our case allows to write:
\begin{align}\label{firstcom}
    D^\dagger a D &= Y +[X,Y] = a +\alpha\qquad
    D^\dagger a^\dagger D = Y +[X,Y] = a^\dagger +\alpha^*.
\end{align}
Further we use:
\begin{equation}
    \frac{d e^{-\beta X(t)}}{dt} = -\int_0^\beta  e^{-(\beta-u) X(t)} \frac{d X(t)}{dt} e^{-uX(t)}du,
\end{equation}
noting that as $d_t X$ generally does not commute with $X$. By proceeding and employing the Campbell identity once again we find:
\begin{equation}\label{dispOP}
     -i\hbar D^\dagger d_t D=i\hbar\left( \frac{d X(t)}{dt}+\frac{1}{2}\left[X(t), \frac{d X(t)}{dt}\right]\right) =i\hbar\left( \dot\alpha^* a-\dot\alpha a^\dagger+\frac{1}{2}(\alpha  \dot\alpha^*-\alpha^* \dot\alpha)\right).
\end{equation} 
By choosing:
\begin{equation}
    \alpha(t)=\alpha_+e^{-i\Omega t}+\alpha_-e^{i\Omega t}= \frac{1}{\hbar}\left[\frac{z_+}{\omega_c-\Omega}e^{-i\Omega t} + \frac{z_-}{\omega_c+\Omega}e^{i\Omega t}\right]
\end{equation}
we arrive to the displaced Hamiltonian
\begin{equation}\label{RotatedH}
    \begin{aligned}
    H'_{\rm LL} &= D^\dagger H_{\rm LL}D -i\hbar D^\dagger d_t D\\&=\hbar\omega_c \bigg(a^\dagger a+\frac{1}{2}\bigg) + \underbrace{\frac{{\bf A}_1(t)^2}{2m}  -\frac{1}{\hbar}\left[\frac{|z_+|^2}{\omega_c-\Omega}+ \frac{ |z_-|^2}{\omega_c+\Omega}+\frac{\omega_c}{\omega^2_c-\Omega^2}( z_+z^*_-e^{-2i\Omega t} +z^*_+z_-e^{2i\Omega t}) \right]}_{\Delta(t)},
\end{aligned}
\end{equation}
which is consistent with our previous study \cite{shi2024floquet}. We can absorb the time dependent constant part of the Hamiltonian into a phase as $\ket{\psi'(t)} = e^{-i \Phi(t)}\ket{\phi(t)}$ with $\Phi(t)=\Phi_0+\int_{t_0}^t \Delta(t^\prime) dt^\prime$. Then the Floquet Landau levels kinetics is mapped onto the standard Landau Level problem, with the full set of solutions labeled by $N,r$, written in Eq.(\ref{state}) of the main text. Notice, since $\Delta(t)$ is independent of Landau level index, the Floquet-Bolzmann equation is unaffected by such phase factor. This can be seen from the fact that in any collision term, the evolution matrix everywhere comes in pair from $t\rightarrow t'$ and $t'\rightarrow t$, thus such $\Delta$-related phases cancel.

\section{Effective cycle averaged free energy and entropy}\label{appB}
\subsection{Floquet Energy}
Here we clarify the definition of the physical energy used in the main text. For a periodically driven non-interacting fermion system with a single-particle Hamiltonian the Floquet states and the steady state density can be written as
\begin{align}
\begin{aligned}
\hat\rho_t &= \sum_{\alpha} f_\alpha \ket{\psi^F_\alpha(t)} \bra{\psi^F_\alpha(t)},\qquad \ket{\psi^F_\alpha(t)} = e^{-i\epsilon^F_\alpha t/\hbar} \sum_{n=-\infty}^{\infty} e^{-in\Omega t} \ket{\phi^n_{\alpha}}
\end{aligned}
\end{align}
where $f_\alpha$ are the steady-state time independent occupations obtained from the Floquet-Boltzmann equation and the instantaneous physical electronic energy is defined from the expectation value of the instantaneous Hamiltonian, $E(t)={\rm Tr}\,[\hat\rho(t)\hat h_t]$.

For transparency, consider first magnetic field free single band driven Floquet problem \cite{matsyshyn2023fermi} with the momentum state label $\ket{\psi^F_\alpha(t)}\equiv\ket{\psi^F_\mathbf{k}(t)}=\exp[-i\int_0^t dt' \epsilon(\mathbf{k}-e\mathbf{A}(t)/\hbar)]\ket{\mathbf{k}}$. Using the Schrodinger equation, we straightforwardly find:
\begin{equation}\label{trueEnergy0}
    \langle E(t)\rangle_T=\langle\sum_\mathbf{k} f_\mathbf{k}\langle \psi^F_\mathbf{k}(t)|\hat h(t)|\psi^F_\mathbf{k}(t)\rangle \rangle_T =\langle\sum_\mathbf{k} f_\mathbf{k}\langle \psi^F_\mathbf{k}(t)|i\hbar\partial_t|\psi^F_\mathbf{k}(t)\rangle \rangle_T= \sum_{\mathbf{k}} f_\mathbf{k}\epsilon_\mathbf{k}^F.
\end{equation}
Thus, in this special zero-field case, the cycle-averaged physical energy equals the occupation-weighted Floquet energy.

For the finite magnetic field Landau level problem, substitution of the Floquet density matrix and employing Eq.~(\ref{state}) of the main text as  
\begin{align}
\begin{aligned}
\ket{\psi^F_\alpha(t)}\equiv\ket{N,r,t} &= e^{-iE_N(t - t_0)/\hbar - (i/\hbar)\int_{t_0}^t \Delta(t')\,dt'}  D[\alpha(t)] \ket{N} \otimes \ket{r},
\end{aligned}
\label{stateSI}
\end{align}

we find the cycle averaged energy as
\begin{equation}\label{trueEnergy}
\begin{aligned}
    \langle E(t)\rangle_T&=\langle\sum_N f_N\langle \psi^F_N(t)|\hat h(t)|\psi^F_N(t)\rangle \rangle_T =\langle\sum_N f_N\langle \psi^F_N(t)|i\hbar\partial_t|\psi^F_N(t)\rangle \rangle_T\\&= \sum_{N} f_N\left(E_N+i\hbar \left\langle \langle N|D^\dagger[\alpha(t)]e^{i\gamma(t)}\partial_t (e^{-i\gamma(t)}D[\alpha(t)])\ket{N} \right\rangle_T \right)
    \\&= \sum_{N} f_N\left(E_N+i\hbar \left\langle -i\dot{\gamma}(t)+\langle N|D^\dagger[\alpha(t)]\dot D[\alpha(t)]|N\rangle\right\rangle_T \right),
\end{aligned}
\end{equation}
where we used the Schrodinger equation in the first line and we decomposed $\Delta(t)=\bar{\Delta}+\delta\Delta(t)$ with $\langle\delta\Delta(t)\rangle_T=0$ [see Eq.~(\ref{RotatedH})] to introduce $\gamma(t) = (1/\hbar)\int_{t_0}^t \delta\Delta(t')\,dt'$. Notice cancellation of $\bar{\Delta}$ as well as $\langle\dot\gamma(t)\rangle_T=0$. Using result derived in Appendix \ref{non_pert} displayed in Eq.~(\ref{dispOP}) we find:
\begin{equation}
    D^\dagger\dot D=\dot{\alpha}a^\dagger-\dot{\alpha}^*a+\frac{1}{2}\left(\alpha^*\dot{\alpha}-\alpha\dot{\alpha}^*\right),\qquad \langle N|D^\dagger\dot D|N\rangle=\frac{1}{2}\left(\alpha^*\dot{\alpha}-\alpha\dot{\alpha}^*\right)
\end{equation}
where we used \(\langle N|a|N\rangle=\langle N|a^\dagger|N\rangle=0\). By explicitly employing $\alpha(t)=\alpha_+e^{-i\Omega t}+\alpha_-e^{i\Omega t}$ [see Appendix \ref{non_pert}] we finally find
\begin{equation}
    \langle E(t)\rangle_T = \sum_{N} f_NE_N+ n_e\hbar \Omega(|\alpha_+|^2-|\alpha_-|^2),
\end{equation}
where $n_e = \sum_N f_N$. Notice, the second term only produces a smooth field-dependent offset in the cycle-averaged energy and does not affect the oscillatory structure discussed in the main text. This justifies using the effective energy $E_{\rm eff}=\sum_N f_NE_N$ in Eq.~(\ref{F_full}).

\subsection{Floquet entropy}
The periodic Gibbs ensemble from Eq.~(\ref{rhot}) of the main text, that characterizes the non-equilibrium steady state, is a non-interacting mixed state of fermions. Namely, in second quantization it should be interpreted as saying that at a given time $t$, the time-dependent single-particle state associated with $\psi^F_\alpha(t)$ in Eq.(\ref{rhot}), is occupied with a probability $f_\alpha$ and empty with probability $1-f_\alpha$, where $f_\alpha$ is the time-independent function that solves the Boltzmann Eq.~(\ref{FBz}). Therefore, interesting, the Von-Neumann entropy of this periodic Gibbs ensemble is a time independent function given by the same elementary formula as for a free fermion gas. We review in this appendix its derivation. For each fermionic mode $\alpha$, there are two possible occupation states, $|0_\alpha,t\rangle, |1_\alpha,t\rangle = c_\alpha^\dagger(t) |0_\alpha,t\rangle$. If the occupation of this mode is $f_\alpha$, then the reduced density matrix of this mode is
\begin{equation}
\hat\rho_\alpha(t)=(1-f_\alpha)\left|0_\alpha;t\right\rangle\left\langle 0_\alpha;t\right|+f_\alpha\left|1_\alpha;t\right\rangle\left\langle 1_\alpha;t\right|.
\end{equation}
For a non-interacting diagonal ensemble, the full many-body density matrix factorizes over fermionic modes:
\begin{equation}
\hat\rho_{\mathrm{MB}}(t)=\prod_\alpha \hat\rho_\alpha(t)=\prod_\alpha\left[(1-f_\alpha)\left|0_\alpha;t\right\rangle\left\langle 0_\alpha;t\right|+f_\alpha\left|1_\alpha;t\right\rangle\left\langle 1_\alpha;t\right|\right].
\label{eq:factorized_many_body_density_matrix}
\end{equation}

Equivalently, one can write the density matrix in the basis of many-body Slater determinants. A many-body configuration is specified by occupation numbers ${\bf n}=\{n_\alpha\},$ with $n_\alpha\in \{0,1\}$ with the corresponding Slater determinant
\begin{equation}
\left|\mathbf n;t\right\rangle=\prod_\alpha\left[\hat c_\alpha^\dagger(t)\right]^{n_\alpha}\left|0\right\rangle.
\end{equation}
Because the modes are independent, the probability of the Slater determinant $|{\bf n},t\rangle$ is $P[{\bf n}]=\prod_\alpha f_\alpha ^{n_\alpha }(1-f_\alpha )^{1-n_\alpha }$. Thus the many-body density matrix can also be written as $\hat\rho_{\rm MB}(t)=\sum_{\bf n}P[{\bf n}]|{\bf n},t\rangle\langle {\bf n},t|$. Which results in the von Neumann entropy of this many-body density matrix to become
\begin{equation}
S=-k_B\mathrm{Tr}[\hat\rho_{\mathrm{MB}}(t)\log\hat\rho_{\mathrm{MB}}(t)]=-k_B\sum_{\mathbf n}P[\mathbf n]\log P[\mathbf n]=-k_B\sum_\alpha\left[f_\alpha\log f_\alpha+(1-f_\alpha)\log(1-f_\alpha)\right],
\label{eq:floquet_entropy_general}
\end{equation}
where we used $ \sum_{\bf n}P[{\bf n}] n_\alpha=f_\alpha,\sum_{\bf n}P[{\bf n}] =1$. Notice, this entropy is independent of time. The Floquet orbitals evolve periodically, but the eigenvalues of the many-body density matrix are the probabilities $P[\mathbf n]$, which are determined only by the time-independent occupations $f_\alpha$.

For the driven Landau-level problem considered in the main text, the generic Floquet label is $\alpha=(N,r)$, where $N$ is the Landau-level index and $r$ is the guiding-center label introduced in Eq.~(7). Magnetic translational symmetry makes the occupation independent of $r$, $f_{N,r}=f_N$. Therefore, for this case,
\begin{equation}
    S=-k_B N_{\phi} \sum_{N}\left[f_N\log f_N+(1-f_N)\log(1-f_N)\right],
\end{equation}
where $N_{\phi}=\sum_{r}1$ is the number of fluxes piercing the 2D plane.

\section{Scattering amplitudes in Landau levels}\label{guiding_center}

In this section, we provide a detailed procedure for performing the summation over the guiding center index. In the undriven Landau level problem, states are labeled as $\ket{\alpha} \equiv \ket{N, r}$, where $N$ is the main Landau level cyclotron index, and $r$ is the guiding center index. Due to magnetic translational invariance, we will show that scattering rates that appear in the kinetic equation can be effectively reduced to depend only on $N$, and the steady state occupation can be taken to only depends on $N$. We also use the equal coupling strength for all the bath modes in the system with $\hat{\chi}_q= \chi_0 \ket{\mathbf{R}}\bra{\mathbf{R}}$, where $\ket{\mathbf{R}}$ is a fermion position eigenket and $\int_{\mathbf{R}} \equiv \int d^d\mathbf{R}/V$, which we introduced in the main text. Following Eq.(\ref{Wet}) from the main text, our  task is to evaluate the Floquet matrix element:
\begin{equation}\label{SIA1}
\int_{\mathbf{R}}\sum_{rr_1}\left|\sum_{l} \bra{\phi^{l}_{N,r}}\chi_{q}\ket{\phi^{l-p}_{N_1,r_1}}\right|^2 = \chi_0^2 \int_{\mathbf{R}} \sum_{rr_1} \left|\sum_{l} \bra{\phi^{l}_{N,r}} \mathbf{R}\rangle\langle \mathbf{R} \ket{\phi^{l-p}_{N_1,r_1}}\right|^2.
\end{equation}
We employ the identity $\ket{\mathbf{R}}\bra{\mathbf{R}} = \delta(\mathbf{R} - \hat{\mathbf{R}}) = (1/A) \sum_q e^{iq(\mathbf{R}- \hat{\mathbf{R}})}$, where the position operator can be decomposed as $\hat{\mathbf{R}} = \hat{\mathbf{R}}_c + \hat{\mathbf{R}}_r$ into orbital, with $\hat{R}^x_c = l_B (\hat a+\hat a^\dagger)/\sqrt{2}, \hat{R}^y_c =- il_B (\hat a-\hat a^\dagger)/\sqrt{2}$, and guiding center index contributions with $\hat{R}^x_m = l_B (\hat b+\hat b^\dagger)/\sqrt{2}, \hat{R}^y_c = il_B (\hat b-\hat b^\dagger)/\sqrt{2}$ where $\hat b~ (\hat b^\dagger)$ are the guiding-center ladder operators. Thus we write:
\begin{align}
   \sum_{rr_1}\left|\sum_{l} \bra{\phi^{l}_{N,r}}\chi_{q}\ket{\phi^{l-p}_{N_1,r_1}}\right|^2&=\chi_0^2\sum_{rr_1}\sum_{ll'} \bra{\phi^{l}_{N,r}}{\mathbf{R}}\rangle \langle{\mathbf{R}}\ket{\phi^{l-p}_{N_1,r_1}}\bra{\phi^{l'-p}_{N_1,r_1}}{\mathbf{R}}\rangle \langle{\mathbf{R}}\ket{\phi^{l'}_{N,r}}\\&=\chi_0^2\sum_{ll'} \sum_{r}\bra{\phi^{l}_{N,r}}{\mathbf{R}}\rangle \langle{\mathbf{R}}\ket{\phi^{l'}_{N,r}}\sum_{r_1}\langle{\mathbf{R}}\ket{\phi^{l-p}_{N_1,r_1}}\bra{\phi^{l'-p}_{N_1,r_1}}{\mathbf{R}}\rangle ,
\end{align}
which can be further simplified by employing:
\begin{align}\label{SIA2}
    \sum_{r}\bra{\phi^{l}_{N,r}}{\mathbf{R}}\rangle \langle{\mathbf{R}}\ket{\phi^{l'}_{N,r}} &= \frac{1}{A}\sum_q \bra{\phi^{l}_{N}}  e^{i q(\mathbf{R}- \hat{\mathbf{R}}_c[a,a^\dagger])}\ket{\phi^{l'}_{N}}\sum_{r}\bra{r}  e^{-i q\hat{\mathbf{R}}_r[b,b^\dagger]}\ket{r} = \frac{ N_\phi}{A}\bra{\phi^{l}_{N}}{\phi^{l'}_{N}}\rangle
\end{align}
where we used $\ket{\phi^{l'}_{N,r}}=\ket{\phi^{l'}_{N}}\otimes \ket{r}$ and the fact that ${\rm Tr}[e^{iq \hat{\mathbf{R}}_r}] = N_\phi \delta_{q,0}$. By substitution of Eq.(\ref{SIA2}) into Eq.(\ref{Wet}) we obtained Eq.(\ref{WetLL}) of the main text.

\section{Evaluation of the trace in  Eq.(\ref{genTR}) of the main text and perturbative limit}\label{trace_eval}

\begin{figure*}
\centering
\includegraphics[width=0.9\textwidth]{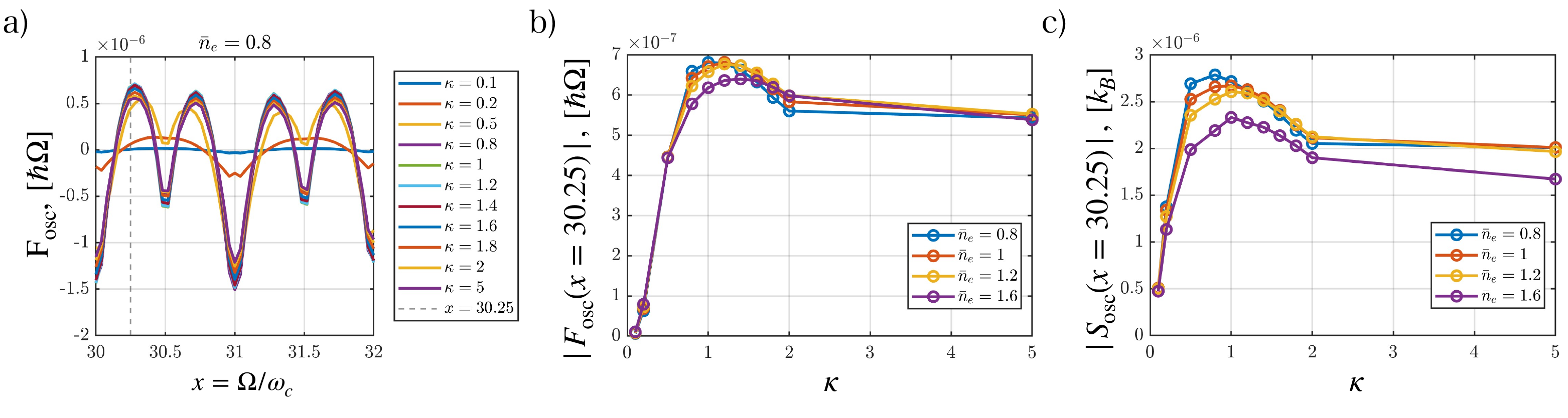}
\caption{(a) Non-perturbative effective free-energy (see Eq.~(\ref{F_full}) of the main text) oscillations at $k_B T_{\rm bath}=0.2\hbar\Omega$, computed with Floquet-harmonic cutoff $p_{\rm max}=2$ in the non-perturbative Floquet-Boltzmann equation, Eq.~(\ref{genEQB}). Amplitudes of the free-energy (b) and entropy (c) oscillations at $\Omega=30.25\omega_c$ as a function of the drive strength $\kappa$, for the effective fillings $\bar n_e$ indicated in the legend. The Floquet oscillations saturate for strong driving, around $\kappa\sim1$.
 }
\label{figSI}
\end{figure*}

Here we show the details of Eq.(\ref{genTR}) derivation in the main text. First, we write by the definition
\begin{equation}
    \sum_{c'}\ket{\phi^{c'}_N}\bra{\phi^{c'-d}_{N_1}} =\int_0^T\int_0^T\frac{dt dt'}{T^2}\left[\sum_{c'}e^{i c'\Omega (t-t')}\right] e^{i d\Omega t'} \ket{N,\alpha(t)}  \bra{N_1,\alpha(t')}
\end{equation}
where $\ket{N,\alpha(t)}\equiv D[\alpha(t)]\ket{N}$and we recognize the Poisson summation formula:
\begin{equation}
    \sum_{c'=-\infty}^\infty e^{i c'\Omega (t-t')}
=\frac{2\pi}{\Omega}\sum_{n=-\infty}^\infty \delta\Bigl(t-t'-\frac{2\pi n}{\Omega}\Bigr)=T\sum_{n=-\infty}^\infty \delta\Bigl(t-t'-nT\Bigr)=T \delta\Bigl(t-t'\Bigr)
\end{equation}
note, we used $-T< t-t' < T$ since $t\in[0,T)$, simplifying to
\begin{equation}
    \sum_{c'}\ket{\phi^{c'}_N}\bra{\phi^{c'-d}_{N_1}} =\int_0^T\frac{dt}{T} e^{i d\Omega t } \ket{N,\alpha(t)}  \bra{N_1,\alpha\left(t\right)}.
\end{equation}

 Next, we employ the orthogonality condition \cite{osti_5185075}:
\begin{equation}
\langle N,\alpha(t') | N,\alpha(t)\rangle = \exp\left\{\frac{1}{2}\Bigl[\alpha(t')^*\alpha(t)-\alpha(t')\alpha(t)^*\Bigr]-\frac{1}{2}|\alpha(t)-\alpha(t')|^2\right\}
L_N\Bigl(|\alpha(t)-\alpha(t')|^2\Bigr).
\end{equation}
and straightforwardly find Eq.(\ref{genTR}) of the main text where
\begin{equation}
    \beta(x,y)= \frac{1}{\hbar}\left[\frac{z_+}{\omega_c-\Omega}(e^{-i2\pi x}-e^{-i2\pi y}) + \frac{z_-}{\omega_c+\Omega}(e^{i2\pi x}-e^{i2\pi y})\right].
\end{equation}

Finally, for the case of the circularly polarized light $z_+$ or $z_-$ vanishes (RHS/LHS light polarization), and we write the simplified Floquet Boltzmann equation as
\begin{gather}
   0=\sum_{N=0}^\infty   \left\{(1-f_{N_1})f_{N} \sum_{p=-\infty}^\infty w_{N,N_1}^pS[\epsilon^F_{N}-\epsilon^F_{N_1}+p\Omega] - (N_1\leftrightarrow N) \right\},\label{genEQB}\\
   S[X] =\nu_B[X](N[X]+1)\Theta[X]+\nu_B[-X]N[-X]\Theta[-X],\\
    w_{N,N_1}^d = \int_{0}^1 du \cos[2\pi d u ]  e^{-\gamma(u)}
L_{N_1}\Bigl[\gamma(u)\Bigr]
L_N\Bigl[\gamma(u)\Bigr], \qquad \gamma(u) =2R_+ (1-\cos [2\pi u]) \label{C8}
\end{gather}
note $w_{N,N_1}^{-d} =w_{N,N_1}^d $and
\begin{equation}
    R_+=\frac{|z_+|^2}{\hbar^2(\omega_c-\Omega)^2}+ \frac{ |z_-|^2}{\hbar^2(\omega_c+\Omega)^2}\approx \frac{|z_+|^2}{\hbar^2\Omega^2}+ \frac{ |z_-|^2}{\hbar^2\Omega^2}
\end{equation}
where $\kappa^2 = R_+\Omega/\omega_c$ is magnetic field independent, positively defined strength of the periodic drive. To see this explicitly employ:
\begin{equation}
    \|z_+\|^2 = {\frac{\hbar\omega_ce^2}{2m\Omega^2}}(\|E_\Omega\|^2+i[\mathbf{E}_\Omega\times\mathbf{E}_{-\Omega}]),\qquad \|z_-\|^2 = {\frac{\hbar\omega_ce^2}{2m\Omega^2}}(\|E_\Omega\|^2-i[\mathbf{E}_\Omega\times\mathbf{E}_{-\Omega}]),
\end{equation}
that follow from Eqs.(\ref{B1}-\ref{B3}).

Critically, in non-perturbative fashion we find saturation of the Floquet oscillations amplitude with amplitude of the drive at $\kappa\sim 1$. The non-perturbative regime is extremely numerically heavy and is out of the scope of current study.

From Eq.(\ref{C8}) we obtain the scattering rate perturbative in the strength of the drive $\kappa$ which is consistent with our previous study \cite{shi2024floquet}:
\begin{equation}
    w_{N,N_1}^d =
\begin{cases}
1 - 2\,(1+N_1+N)\,\Bigl(|\alpha_+|^2+|\alpha_-|^2\Bigr) + O(\mathcal{E}^4), & d=0, \\[1mm]
\,(1+N_1+N)\,\Bigl(|\alpha_+|^2+|\alpha_-|^2\Bigr) + O(\mathcal{E}^4), & d=\pm1, \\[1mm]
0 + O(\mathcal{E}^4), & |d|\ge2.
\end{cases}
\end{equation}

\begin{figure*}
\centering
\includegraphics[width=0.6\textwidth]{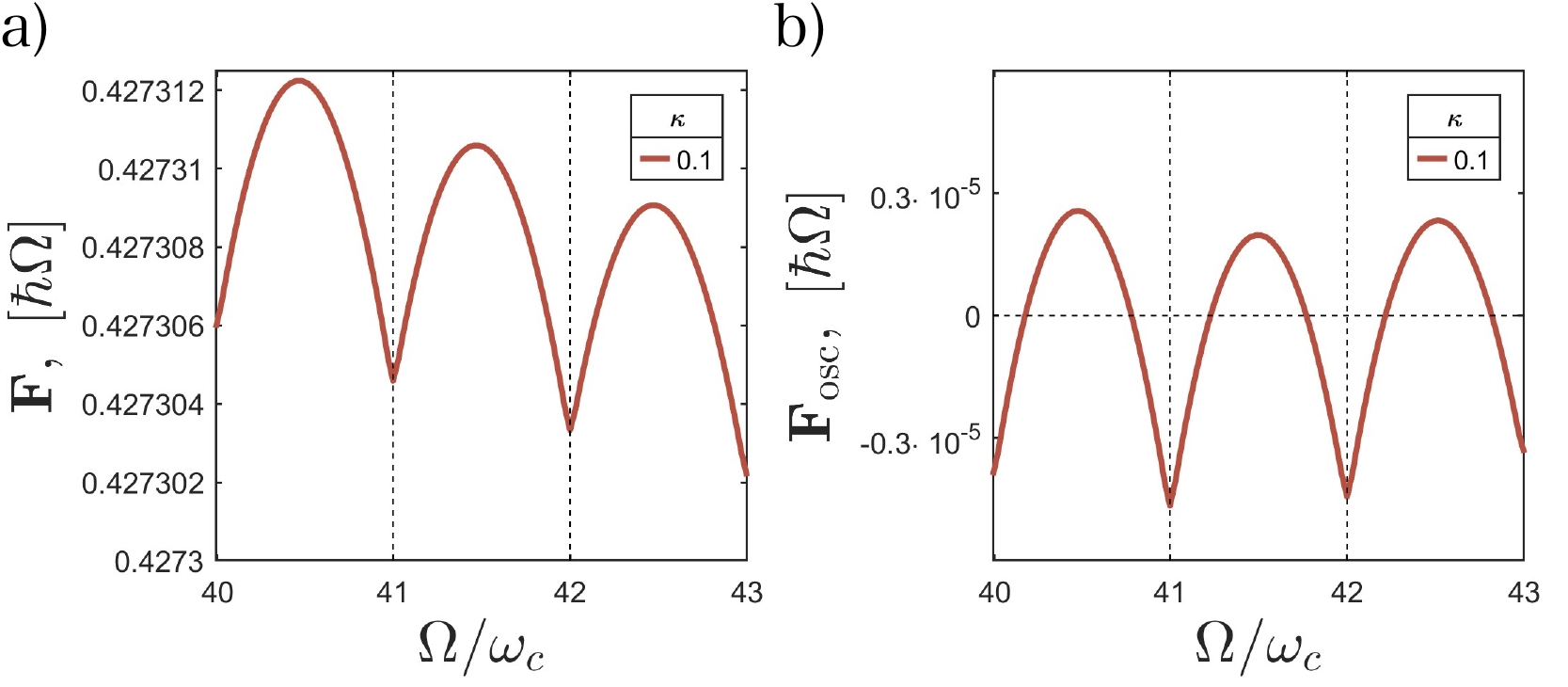}
\caption{(a) Effective free energy, defined in Eq.~(\ref{F_full}) of the main text, as a function of magnetic field, parametrized by $\omega_c$. (b) Corresponding isolated oscillatory component. }
\label{figSI2}
\end{figure*}

\section{Oscillation extraction}\label{Oscillation_extraction}

In this section we discuss the numerical calculation of the Floquet free energy oscillations introduced in Eq.(\ref{F_full}) of the main text. We begin by fixing of the total particle density for each magnetic field. In 2D electron gas at $T=0$, the electron density is related to the Fermi energy as $\mu = 2\pi \hbar^2 n_e/m$, while the Landau level degeneracy per area is $N_\phi/A = e B/h = eB/(2\pi\hbar)$ with the cyclotron frequency $\omega_c = eB/m$. Thus we impose the normalization:
\begin{equation}
    \lim_{B\rightarrow0}\frac{N_\phi}{A} \sum_N f(N) = n_e = \frac{m \mu}{\pi \hbar^2}=\left[\frac{N_\phi}{A}\right]\frac{m \mu A}{\pi \hbar^2N_\phi}=\left[\frac{N_\phi}{A}\right]\frac{m \mu }{ \hbar eB}=\left[\frac{N_\phi}{A}\right]\frac{ \mu }{ \hbar \omega_c}.
\end{equation}

The quantity for which we compute the oscillations is the following effective free energy:
\begin{equation}
    F = E - T_{\rm bath} S = E -  T_{\rm bath} N_\phi S,\qquad \frac{F}{n_e A\hbar\Omega}=\frac{\omega^2_c}{\Omega^2\bar\mu}\sum_N \left( f(\epsilon_N)(N+1/2)-\frac{T}{\omega_c}S_N\right) .
\end{equation}
To extract the strictly oscillatory part, we further subtract the non-oscillatory (Landau diamagnetic) component to identify the quantum oscillations see Fig.\ref{figSI2}.

\section{ohmic Bath Quantum oscillations}\label{ohmic_bath_osc}

\begin{figure*}[t]
\centering
\includegraphics[width=0.6\textwidth]{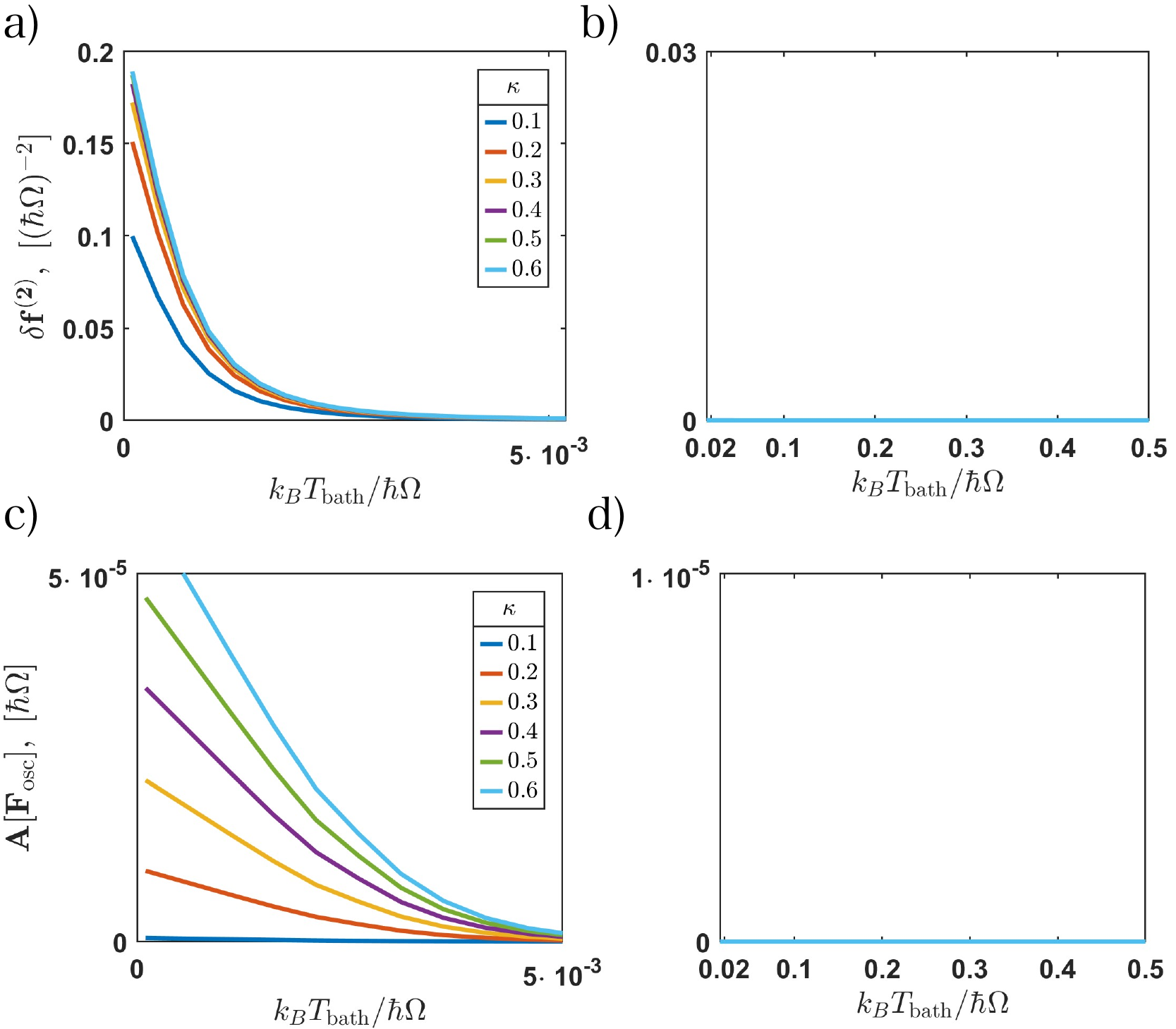}
\caption{Jump of the second derivative of the occupation function at $k=k_F$ for a perfectly ohmic bath, $c_2=0$, as a function of bath temperature and for several drive strengths $\kappa$ in (a,b). Corresponding quantum-oscillation amplitude in the interval $\Omega/\omega_c\in[40,43]$ in (c,d). We use $\bar n_e=0.8$. In contrast to Fig.~\ref{fig3} of the main text, the high-temperature revival is absent for the perfectly ohmic bath.
}
\label{figSIE6}
\end{figure*}

Here we show that for a perfectly ohmic bath (i.e $c_2$=0), the “high temperature oscillations” associated with revival are absent. This is consistent with the results of  Ref\cite{PhysRevLett.134.196401}, which found that the second derivative discontinuity for the perfectly Ohmic bath is only sharp when the bath is at zero temperature and it is smeared out when the bath is at finite temperature. In agreement with this, we find that the quantum oscillations only are present in the low temperature regime in this case, as shown in Fig.\ref{figSIE6}.

\section{Oscillations Floquet regime phase shift}\label{MOKE}
\begin{figure*}
\centering
\includegraphics[width=0.6\textwidth]{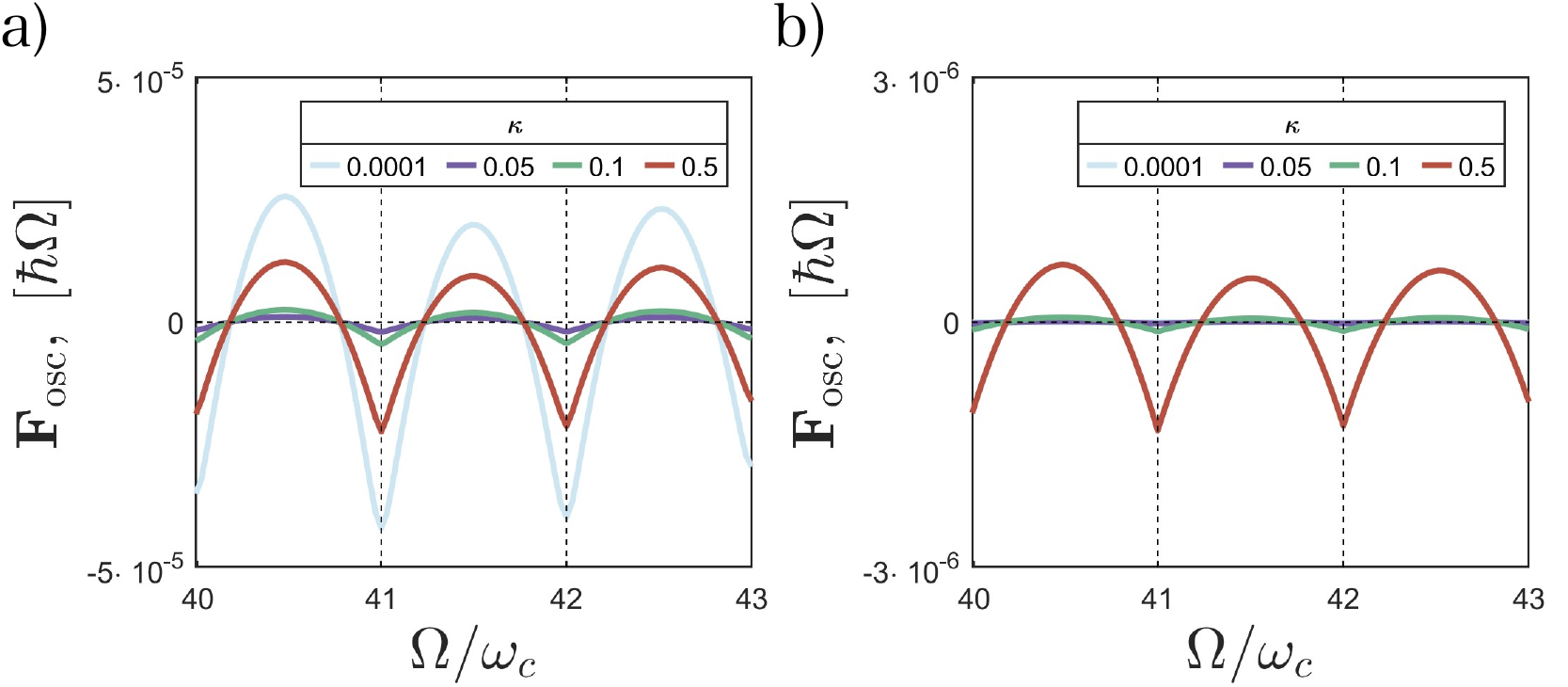}
\caption{(a) Low-temperature quantum oscillations at $k_B T_{\rm bath}=10^{-4}\hbar\Omega$. We find that the oscillation phase remains unchanged across the crossover from the conventional equilibrium regime (blue line) to the Floquet regime (red line). (b) High-temperature regime at $k_B T_{\rm bath}=0.1\hbar\Omega$. The equilibrium oscillations are exponentially suppressed and remain close to the $F_{\rm osc}=0$ line, while the Floquet-controlled component grows with increasing drive strength. Here we choose $\bar n_e=1$, so that the equilibrium Fermi energy is equal to $\hbar\Omega$.}
\label{figSI7}
\end{figure*}

We contrast the oscillation phases in the deep Floquet regime to the equilibrium quantum oscillations. To do this we choose the electron density and the driving frequency to have a value such that the area of the usual equilibrium Fermi surface and the Floquet Fermi surface are identical, and thus their quantum oscillations frequency is the same. The idea is that this allows us to contrast more easily their phase to determine if there are any changes. We find, however, that the phase as a function of magnetic field is preserved during the crossover from usual oscillations which occurs as the driving amplitude, $\kappa$, increases, as shown in FIG.\ref{figSI7}, which illustrates that minima and maxima do not shift as the amplitude of the drive, $\kappa$ increases. Therefore, the phase of these oscillations is not consistent with the one observed in the regime of coexistence of MIRO and SdH oscillations reported in Ref.\cite{shi2015shubnikov}.

\section{Fit parameters for the low and high temperature discontinuity behavior}\label{fit_param}
Here we present values of the parameter fit to for the low and high temperature occupation function discontinuity fits, see Table \ref{fitparam}.
\begin{table}[h]
\centering
\begin{tabular}{|c||c|c|c||c|c|}
\hline
\multicolumn{1}{|c||}{} & \multicolumn{3}{c||}{\textbf{Low temperature fit}} & \multicolumn{2}{c|}{\textbf{Higher temperature fit}} \\
\hline\hline
$\kappa$ & $a$ & $b$ & $c$ & $a'$ & $b'$ \\
\hline\hline
0.1 & 0.097 & 0.3  & $4.4\times10^{-4}$ & $9\times10^{-4}$ & $8\times10^{-3}$ \\\hline
0.3 & 0.15  & -0.7 & $4.4\times10^{-4}$ & $6\times10^{-3}$ & $3.6\times10^{-2}$ \\\hline
0.4 & 0.165  & 1.06 & $4.4\times10^{-4}$ & $1.6\times10^{-2}$ & $0.1$ \\
\hline
\end{tabular}
\caption{Fitted parameters for the jump in the second derivative of the occupation function. In the low-temperature regime, region I, the jump is fitted by $\delta f^{(2)}[\tau\to0]=(a+b\tau)e^{-\tau/c}$. In the high-temperature regime, region II, it is fitted by $\delta f^{(2)}[\tau\gg\tau_{\rm max}]=a'\tau/(b'+\tau^2)$. For a comparison between the fits and the numerically simulated data, see Fig.~\ref{fig3}c.}
\label{fitparam}
\end{table}

\end{document}